\documentclass[
prb,
reprint,
superscriptaddress,
showpacs,
amsmath,amssymb,
floatfix,
]{revtex4-1}
\usepackage{dcolumn}
\usepackage{bm}
\usepackage{turnstile}

\usepackage{graphicx}
\usepackage{amsmath,amssymb}
\usepackage{url}
\usepackage{multirow}
\usepackage{float}
\usepackage[toc,page]{appendix}

\DeclareMathOperator{\sgn}{\mathop{\mathrm{sgn}}}
\DeclareMathOperator{\re}{\mathop{\mathrm{Re}}}

\newcommand{\Eq}[1]{Eq.~(\ref{#1})}
\newcommand{\Eqs}[1]{Eqs.~(\ref{#1})}
\begin{document}

	\title{Reentrant superconductivity in proximity to a topological insulator}
	\author{T. ~Karabassov}
	\affiliation{HSE University, 101000 Moscow, Russia}
	\date{\today}
	\author{A.~A.~Golubov}
	\affiliation{Faculty of Science and Technology and MESA$^+$ Institute for Nanotechnology,
		University of Twente, 7500 AE Enschede, The Netherlands}
	\affiliation{Moscow Institute of Physics and Technology, 141700 Dolgoprudny, Russia}
	\author{V.~M.~Silkin}
	\affiliation{Donostia International Physics Center (DIPC), Paseo Manuel de Lardizabal 4, San Sebasti\'{a}n/Donostia, 20018 Basque Country, Spain}
	\affiliation{Departamento de F\'{\i}sica de Materiales, Facultad de Ciencias Qu\'{\i}micas,
		UPV/EHU, 20080 San Sebasti\'{a}n, Basque Country, Spain}
	\affiliation{IKERBASQUE, Basque Foundation for Science, 48011 Bilbao, Spain}
	\author{V.~S.~Stolyarov}
	\affiliation{Moscow Institute of Physics and Technology, 141700 Dolgoprudny, Russia}
	\affiliation{Dukhov Research Institute of Automatics (VNIIA), 127055 Moscow,  Russia}
	\author{A.~S.~Vasenko}
	\email{avasenko@hse.ru}
	\affiliation{HSE University, 101000 Moscow, Russia}
	\affiliation{I.E. Tamm Department of Theoretical Physics, P.N. Lebedev Physical Institute, Russian Academy of Sciences, 119991 Moscow, Russia}
	\begin{abstract}
		In the following paper we investigate the critical temperature $T_c$ behavior in the two-dimensional S/TI (S denotes superconductor and TI - topological insulator) junction with a proximity induced in-plane helical magnetization in the TI surface. The calculations of $T_c$ are performed using the general self-consistent approach based on the Usadel equations in Matsubara Green's functions technique. We show that the presence of the helical magnetization leads to the nonmonotonic behavior of the critical temperature as a function of the topological insulator layer thickness.
	\end{abstract}
	
	
	\maketitle
	
	\section{Introduction}
	
	%
	Topological state of matter has been receiving a lot of attention for the past decade.\cite{Fu2007, Hasan2010, Sato2017, book1, book2} Particularly, three-dimensional topological insulators (3D TI) have large potential for fault tolerant quantum computation.\cite{Sarma2006, Aguado2020} This is possible due to strong spin-orbit coupling (SOC) and time reversal symmetry that take place in such materials. There are special topologically protected states on the surface of the 3D topological insulator. These surface states are Dirac helical states, i.e., their spin and momentum are coupled in a well defined way resulting in a spin-momentum locking effect. Interesting transport properties are revealed when topological insulator and superconductor are in proximity to each other, forming a hybrid structure.\cite{Qi2011} In such hybrids in the presence of a magnetic field or a magnetic moment of an adjacent ferromagnet, zero energy Majorana modes can arise. \cite{Fu2008, Tanaka2009, Sato2009, Alicea2012, Beenakker2013, Tkachov2013}

	The proximity effect \cite{Demler1997, Ozaeta2012R, Bergeret2013, Bobkova2017, BergeretRMP, BuzdinRMP} that takes place in superconducting hybrid structures can lead to various phenomena occurring near interfaces. For instance, critical temperature $T_c$ behaves nonmonotonically as a function of different system parameters in S/F bilayers with uniform magnetization \cite{Fominov2002} and multilayered S/F spin-valves with a magnetization misalignment in F layers\cite{Fominov2010}. Particularly, $T_c$ demonstrates reentrant behavior under certain range of parameters.\cite{Fominov2002} Such behavior originates from nontrivial dependence of Cooper pair wavefunction, which can also result in oscillating Josephson critical current \cite{Buzdin1982, Buzdin1991, Ryazanov2001, Oboznov2006, Vedyayev2006, Vasenko2008, Bakurskiy2017}, density of states \cite{Kontos2001, Vasenko2011} and critical temperature \cite{Jiang1995, Tagirov1998, Proshin2001, Khaydukov2018, Karabassov2019} in S/F/S junctions.
	
	According to the theory developed in Refs.~\onlinecite{Tanaka2009, Zuyzin2016}, there are no Josephson critical current oscillations in hybrid S/TI/S structures with a proximity induced uniform in-plane field in the TI layer. At the same time, as predicted in Ref.~\onlinecite{Zuyzin2016}, the   critical current demonstrates oscillatory behavior in case when the TI surface with helical magnetization serves as a weak link.
	Following the results of Zuyzin {\it et al.} \cite{Zuyzin2016}, the observation of $0-\pi$ phase transitions in the critical supercurrent may imply nontrivial critical temperature behavior and in particular the reentrant $T_c$ behavior in the S/TI junction. Therefore, investigation of the $T_c$ in the hybrids with both spin-orbit coupling and helical magnetization is essential for further understanding of the underlying physics and potential future applications. As far as we know, the critical temperature in S/TI structures has not been studied yet.
	
	Spin-orbit effects have been discussed actively in the framework of the quasiclassical Green's functions approach in layered structures.\cite{Bergeret2013, Bergeret2014, Jacobsen2015,Bujnowski2019,Eskilt2019, Nashaat2019,Bobkova2016, Bobkova2017,Alidoust2018, Alidoust2020} Recently, the generalized quasiclassical theory was developed for a two-dimensional system with SOC and an exchange field both much greater than the disorder strength.\cite{Lu2019} It has been shown that spontaneous supercurrent can flow in a Josephson junction, where magnetized superconductors are weakly coupled through the surface of 3D TI.\cite{Alidoust2017} 
	
	The goal of this work is to provide a quantitative investigation of the critical temperature in the S/TI hybrid structure as a function of its parameters applying the quasiclassical Green's function approach. The helical magnetization pattern considered in this paper is similar to one previously studied in S/F bilayers with nonuniform spiral magnets. \cite{Champel2005, Champel2005_2, Champel2008} Particularly, the superconducting spin valve consisting of a superconducting layer and a spiral magnetic was proposed for the spintronic application, using re-orientation of the spiral direction as a method of the spin-valve control. \cite{Pugach2017,Pugach2017_2, Pugach2018}	However, nature of the effects that appear in our structure is different, since they are caused not only by in-plane helical magnetization pattern, but also by the spin-orbit coupling. 
	
	The paper is organized as follows. In Sec.~\ref{Secmod}, we formulate the theoretical model and basic equations for the cases of $h(y)$ and $h(x)$ helical magnetizations. In Sec.~\ref{Results} we present the results of the critical temperature calculations using the single-mode approximation. The results are concluded in Sec.~\ref{Conclusion}.
	
	\section{Model}\label{Secmod}
	\begin{figure*}[t]
		\centering
		\includegraphics[width=2.0\columnwidth]{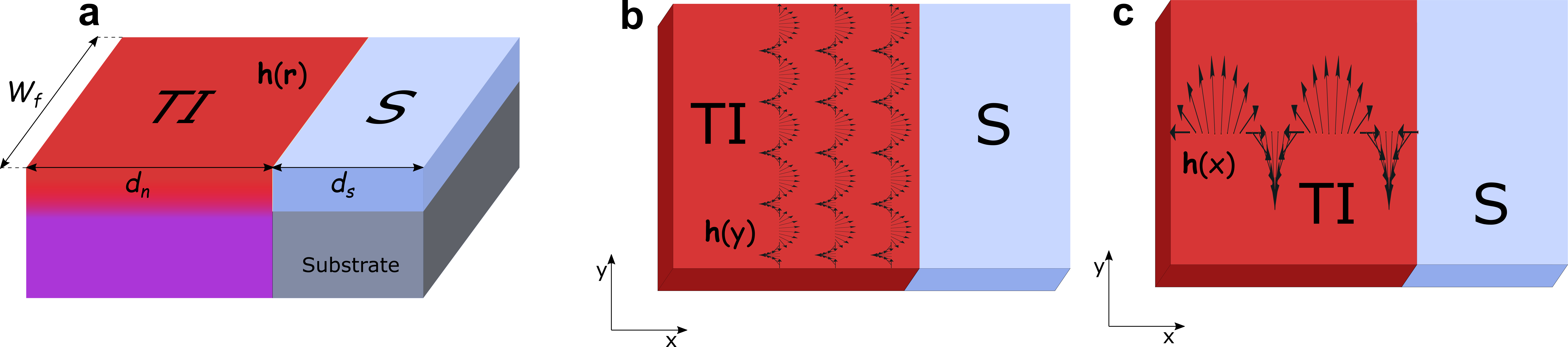}
		\caption
		{
			(Color online) (a) Schematic of a 3D topological insulator (TI) - superconductor (S) junction with a proximity induced helical magnetization pattern. The magnetization vector is given by $\textbf{h}(y)=h_0(\cos Qy, \sin Qy, 0)$ (b) and $\textbf{h}(x)=h_0(\cos Qx, \sin Qx, 0)$ (c). The junction resides in the $\textit{x-y}$ plane and the S/TI interface lie in the $y$ direction at $x=0$. $d_n$ and $d_s$ are the thicknesses of TI and S layers respectively, while $W_f$ is the width of the junction. $\gamma_B$ is a transparency parameter which is proportional to the interface resistance.}
		\label{Fig_Model}
	\end{figure*}
   In this work we consider the 2D nanostructure, which is depicted in the Fig.~\ref{Fig_Model}. It consists of superconductor S of thickness $d_s$ and topological insulator (TI) of thickness $d_n$ with a proximity induced helical magnetization pattern of the following types:
		\begin{align} \label{h_y}
		\textbf{h}(y)=h_0(\cos Qy, \sin Qy, 0), \\
		\textbf{h}(x)=h_0(\cos Qx, \sin Qx, 0),\label{h_x}
		\end{align}
	where $Q=2\pi/ \lambda$ and $\lambda$ determines the actual pattern of helical magnetization. It is important to note that we consider the variations of the magnetization $\textbf{h}$ in the $\textit{x-y}$ plane. Similar helical pattern with a period $\lambda \approx 10$nm was observed experimentally in manganese on a tungsten substrate.\cite{Bode2007} The orientation of the structure is along the $\textit{x}$ direction. In order to observe the inverse proximity effect the superconductor must be two-dimensional. Such disordered homogeneous superconducting 2D films can be obtained with the help of modern deposition techniques \cite{Brun2016}.
	
	To calculate the critical temperature $T_c(d_n)$ of this structure we assume the diffusive limit, when the elastic scattering length $\ell$ is much smaller than the
	coherence length, and use the framework of the linearized Usadel equations for the S and TI layers in Matsubara representation.\cite{Belzig1999, Usadel1970}. We perform the calculations in the low proximity limit expanding the Green's function around the bulk solution,
	\begin{equation}
	\hat{g}=
	\begin{pmatrix}
	\sgn\omega_n & f \\
	-\bar{f} & -\sgn\omega_n .
	\end{pmatrix}
	\end{equation}
	Such limit is experimentally feasible and can be easily achieved in the vicinity of the superconducting critical temperature $T_c$ or in a hybrid structures with low transparent interfaces.
	\subsection{Helical magnetization h(y)}
	
	In this subsection we establish the equations for the magnetization pattern evolving along the S/TI interface indicated in Eq.~\eqref{h_y}, i. e. in $y$ direction. Since the low proximity limit is assumed, near $T_c$ the normal Green's function in a superconductor is $g_s = \sgn \omega_n$, and the Usadel equation for the anomalous Green's function $f_s$ take the following form. In the S layers ($0 < x <d_s$) it reads 
	\begin{equation}\label{Usadel_S}
	\xi_s^2 \pi T_{cs} \left(\frac{\partial^2}{\partial x^2}+\frac{\partial^2}{\partial y^2}\right)f_s - |\omega_n| f_s + \Delta = 0.
	\end{equation}
	In the TI layer we consider the Usadel equation derived in  Ref.~\onlinecite{Zuyzin2016}, 
	\begin{equation}\label{Usadel_TI}
	\left(\frac{\partial}{\partial x}-\frac{2i}{\alpha} h_y(y)\right)^2 f_T + \left(\frac{\partial}{\partial y}+\frac{2i}{\alpha} h_x(y)\right)^2 f_T = \frac{|{\omega_n}|}{\xi_n^2 \pi T_{cs}} f_T
	\end{equation}
	Since we consider the dirty limit, the spinless Green's function matrix $\hat{g}_s$ is used in our calculations, whereas the spin texture is contained in the matrix $ \check{g}(\textbf{n}_F)=\hat{g} \left(1+ \hat{\eta} \cdot \textbf{n}_F \right)/2$, where $\textbf{n}_F=\textbf{p}_F / p_F$ and $\hat{\eta}= \left(-\sigma_2, \sigma_1\right)$. The spin-momentum locking effect can be seen from the spin matrix $\check{g}$, so that spin and momentum are always fixed at the right angle.
	
	Finally, the self-consistency equation reads,\cite{Belzig1999}
	\begin{equation}\label{Delta}
	\Delta \ln \frac{T_{cs}}{T} = \pi T \sum_{\omega_n} \left ( \frac{\Delta}{|\omega_n|} - f_s \right ).
	\end{equation}
	In \Eqs{Usadel_S}-\eqref{Delta} $\xi_s = \sqrt{D_s/ 2\pi T_{cs}}$, $\xi_n = \sqrt{D_n/ 2\pi T_{cs}}$, $\omega_n = 2 \pi T (n + \frac{1}{2})$, where $n = 0, \pm 1, \pm 2, \ldots$ are the Matsubara frequencies, $T_{cs}$ is the critical temperature of the superconductor S, and $f_{s(T)}$ denotes the singlet components of anomalous Green function in the S(TI) region (we assume $\hbar = k_B = 1$).
	
	As far as our 2D system is periodic in $y$ direction and large values of helical magnetization parameter $Q$ are considered such that $\lambda \ll W_f$, we can expand the anomalous Green functions using the Fourier series. The function $f_T$ then can be written as,
	\begin{align}
	f_T(x,y)=\sum_{p=-\infty}^{+\infty} f^{(p)}_T(x) e^{i p Q y}.
	\end{align}
	The Usadel equation in the TI layer for the amplitudes $f_T^{(p)}$ then takes the following form,
	\begin{align}\label{Usadel_TI_Fourier}
	\left(\frac{\partial}{\partial x}-\frac{2i}{\alpha} h_y(y)\right)^2 f_T^{(p)} - p^2 Q^2 f_T^{(p)} - \frac{4 p Q h_x(y)}{\alpha} f_T^{(p)} &= \nonumber \\ \left( \frac{|{\omega_n}|}{\xi_n^2 \pi T_{cs}} + \frac{4 h_x^2(y)}{\alpha^2} - \frac{2 i h'_x(y) }{\alpha}\right) f_T^{(p)},
	\end{align}
	where, $h'_x(y)$ is a derivative of $h_x$ along the $y$ direction.
	Whereas, in the S layer the singlet function $f_s$ as well as $\Delta$ can also be expanded into a Fourier series,
	\begin{align}
	f_s(x,y)=\sum_{p=-\infty}^{+\infty} f^{(p)}_s(x) e^{i p Q y},\\
	\Delta(x,y)=\sum_{p=-\infty}^{+\infty} \Delta^{(p)} (x) e^{i p Q y}. 
	\end{align}
	The amplitudes $f^{(p)}_s$ obey the following equation,
	\begin{align}
	\xi_s^2 \left(\frac{\partial^2 f^{(p)}_s}{\partial x^2} - p^2 Q^2 f^{(p)}_s -\frac{|\omega_n|}{\xi_s^2 \pi T_{cs}} f^{(p)}_s \right) +  \frac{\Delta^{(p)}}{\pi T_{cs}} = 0.
	\end{align}
	The self-consistency equation for the Fourier amplitudes in the superconductor can be written as,
	\begin{align}
	\Delta^{(p)} \ln \frac{T_{cs}}{T} = \pi T \sum_{\omega_n} \left ( \frac{\Delta^{(p)}}{|\omega_n|} - f^{(p)}_s \right ).
	\end{align}
	From the equations above, it is clear that the amplitudes of the Fourier series are decoupled in the vicinity of the critical temperature. Therefore, each Fourier component $p$ satisfies certain Usadel equation and the boundary conditions. Moreover, every single Fourier harmonic $p$ of anomalous Green function $f^{(p)}_s$ and pair amplitude $\Delta^{(p)}$ determines particular $T_c$ through the corresponding gap equation. However, the physical solution is the one, which gives the highest critical temperature $T_c$, i. e. the solution is energetically favorable.
	
	We also need to supplement the equations above with proper boundary conditions to solve the problem.\cite{KL,Zuyzin2016} We assume low transparency limit of the interface between topological insulator (TI) and superconducting layer (S). It is also assumed that spin is conserved when the electrons tunnel across the interface, whereas momentum is not conserved.
	For the Fourier harmonics of the solution $f^{(p)}$ that we have introduced above taking all the simple transformations into account, the boundary conditions at $x=0$ take the form,
	\begin{equation}\label{KL1}
	\gamma_B \xi_n \left(\frac{\partial}{\partial x} - \frac{2 i h_y(y)}{\alpha}\right) f_T(0) = f_s(0) - f_T(0) ,
	\end{equation} 
	\begin{equation}\label{KL2}
	\gamma \xi_n \left(\frac{\partial}{\partial x} - \frac{2 i h_y(y)}{\alpha}\right) f_T(0)= \xi_s \frac{\partial f_s(0)}{\partial x}.
	\end{equation} 
	Here we omitted the component index $p$. The parameter $\gamma_B=R_b \sigma_n / \xi_n$ is transparency parameter which is the ratio of resistance per unit area of the surface of the tunneling barrier to the resistivity of the TI layer and describes the effect of the interface barrier \cite{KL, gamma_b}. In \eqref{KL2} the dimensionless parameter $\gamma = \xi_s \sigma_n / \xi_n \sigma_s$ determines the strength of suppression of superconductivity in the S layers near the S/TI interface compared to the bulk (inverse proximity effect).
	No suppression occurs for $\gamma = 0$, while strong suppression takes place for $\gamma \ll 1$. Here $\sigma_{s(n)}$ is the normal-state conductivity of the S(TI) layer.
	These boundary conditions should also be supplemented with vacuum conditions at the edges ($x=-d_n$ and $x=+d_s$),
	\begin{align}\label{Vacuum}
	\frac{\partial f_s(d_s)}{\partial x}=0, \quad
	\left(\frac{\partial}{\partial x} - \frac{2 i h_y(y)}{\alpha}\right) f_T(-d_n) =0.
	\end{align}

	The solution of the equation \eqref{Usadel_TI_Fourier} can be found in the form,
	\begin{equation}\label{f_T}
	f^{(p)}_T= C(\omega_n) \cosh \kappa_{p,y} \left( x + d_n \right) \exp{\left[ i \frac{2 h_y(y)}{\alpha} (x + d_n)\right]},
	\end{equation}
	where,
	\begin{align*}
	k_{p,y} &= \sqrt{\frac{|{\omega_n}|}{\xi_n^2 \pi T_{cs}} + \frac{4}{\alpha^2}  h_x^2(y) - \frac{2 i h'_x(y) }{\alpha} + Q_p}, \\
	Q_p &= p^2 Q^2 + \frac{4 p Q h_x(y)}{\alpha}.
	\end{align*}
	Here $C(\omega_n)$ is the coefficient, which is found from the boundary conditions and the wavevector acquires additional imaginary term due to fast oscillations of the anomalous Green's function along the $y$ direction compared to the case of uniform magnetization ($Q=0$). The introduced solution to the equation automatically satisfies the vacuum boundary conditions \eqref{Vacuum}.
	
	As far as $\Delta$ is assumed to be real valued function, we write our equations for anomalous Green's functions in real form. Also we consider only positive Matsubara frequencies $\omega_n$.
	Following the standard procedure we obtain final set of equations which are sufficient to calculate critical temperature as a function of $d_n$.
	
	Using the boundary conditions \eqref{KL1}-\eqref{KL2} we would like to write the problem in a closed form with respect to the Green's function $f_s$. At $x=0$ the boundary conditions can be written as:
	\begin{equation}\label{boundary}
	\xi_s \frac{\partial f_s(0)}{\partial x} = \frac{\gamma}{\gamma_b + A_{pT} (\omega_n)} f_s(0),
	\end{equation}
	where
	\begin{align*}
	A_{pT}(\omega_n)= \frac{1}{k_{p,y}} \coth{ k_{p,y} d_n}.
	\end{align*}
	
	The boundary condition \eqref{boundary} is complex. In order to rewrite it in a real form, we use the following relation,
	\begin{equation}\label{symformula}
	f^\pm = f(\omega_n) \pm f(-\omega_n).
	\end{equation}
	According to the Usadel equation \eqref{Usadel_S}, there is a symmetry relation $f(-\omega_n) = f^*(\omega_n)$ which implies that
	$f^+$ is a real while $f^-$ is a purely imaginary function.
	Then, we rewrite the Usadel equation in the S layer in terms of $f_s^+$ and $f_s^-$ utilizing symmetry relation \eqref{symformula}. Since the pair potential $\Delta$ is considered to be real valued function, we can find the solution analytically in the Usadel equation for the imaginary function $f_s^-$. Using the solution found analytically, it is possible to derive the complex boundary condition \eqref{boundary} in real form for the function $f_s^+$,
	\begin{equation}\label{B1a}
	\xi_s \frac{\partial f^+_s(0)}{\partial x} = W^{(p)}(\omega_n) f^+_s(0),
	\end{equation}
	where we used the notations,
	\begin{align}
	W^{(p)}(\omega_n) &= \gamma \frac{A_{ps} \left (\gamma_b + \re A_{pT} \right ) + \gamma}{A_{ps} |\gamma_b + A_{pT}|^2 + \gamma (\gamma_b + \re A_{pT})},
	\\
	A_{ps} &= \kappa_{ps} \tanh \kappa_{ps} d_s, \quad A_{pT}(\omega_n)= \frac{1}{k_{p,y}} \coth{ k_{p,y} d_n},\nonumber
	\\
	\kappa_{ps} &=\sqrt{Q^2 p^2 +\frac{|{\omega_n}|}{\xi_s^2 \pi T_{cs}}}. \nonumber
	\end{align}

	In the same way we rewrite the self-consistency equation for $\Delta$ in terms of symmetric function $f_s^+$ considering only positive Matsubara frequencies,
	\begin{equation}\label{Delta+}
	\Delta^{(p)} \ln \frac{T_{cs}}{T} = \pi T \sum_{\omega_n > 0} \left ( \frac{2\Delta^{(p)}}{\omega_n} - f_s^{(p)+} \right),
	\end{equation}
	as well as the Usadel equation in the superconducting S layer,
	\begin{equation}\label{finUsadelS}
	\xi_s^2 \left(\frac{\partial^2 f^{(p)+}_s}{\partial x^2} - \kappa_{ps}^2 f^{(p)+}_s\right) + \frac{2 \Delta^{(p)}}{\pi T_{cs}} = 0.
	\end{equation}
	To calculate the critical temperature in the system considered, we use the equations \eqref{B1a}-\eqref{finUsadelS}, together with the vacuum boundary condition \eqref{Vacuum} for the Fourier components $f^{(p)+}_s$.
	
	\subsection{Helical magnetization h(x)}
	Here we consider the system consisting of a superconductor and topological insulator with helical magnetization pattern presented in the Eq.~\eqref{h_x}. In this case the Usadel equation should be rewritten in terms of magnetization $\textbf{h}(x)$. We assume that the anomalous Green's function does not depend on $y$ coordinate and thus the corresponding derivatives are neglected. The Usadel equation in the TI layer then takes the following form,
	\begin{equation}\label{Usadel_TI_x}
	 \left(\frac{\partial}{\partial x} - \frac{2 i h_y(x)}{\alpha}\right)^2 f - \frac{4 h_x^2(x) }{\alpha^2}  f = \frac{|{\omega_n}|}{\xi_n^2 \pi T_{cs}} f
	\end{equation}
	In order to rewrite the Eq.~\eqref{Usadel_TI_x} in real form we introduce the following anzatz,
	\begin{equation}
	f(x)= f_L(x) \exp \left[ - i \frac{2 h_0}{\alpha Q} \cos Q x\right].
	\end{equation}
	Inserting this substitution into the Eq.~\eqref{Usadel_TI_x}, we obtain the equation for real valued function in the TI layer,
	\begin{equation}\label{Usadel_TI_re}
	 \frac{\partial^2 f_L}{\partial x^2} = \left(\frac{|{\omega_n}|}{ \xi_n^2 \pi T_{cs}} +  \frac{4 h_0^2 \cos^2 Q x }{\alpha^2} \right) f_L
	\end{equation}
	For this system we utilize the same boundary conditions as in previous subsection and express them in real form using symmetry relation \eqref{symformula}. After the substitutions the boundary conditions take the form,
	\begin{equation}\label{KL1_x}
	\gamma_B \xi_n \frac{\partial f_L(0)}{\partial x} = C_0 f^+_s(0) - f_L(0) ,
	\end{equation}
	\begin{equation}\label{KL2_x}
	\gamma \xi_n \frac{\partial f_L(0)}{\partial x} = \xi_s C_0 \frac{\partial f^+_s(0)}{\partial x}.
	\end{equation}
	where $C_0= \cos \left( 2 h_0 / \alpha Q\right)$. Finally, the boundary conditions at the free edges at $x=d_s$ and $x=-d_n$,
	\begin{align}\label{Vacuum_x}
	\frac{\partial f_s(d_s)}{\partial x}=0, \quad
	\frac{\partial f_T(-d_n) }{\partial x} =0.
	\end{align}

	Similarly, we introduce the self-consistency equation for $\Delta$ in terms of symmetric function $f_s^+$ treating only positive Matsubara frequencies,
	\begin{equation}\label{Delta+_x}
	\Delta \ln \frac{T_{cs}}{T} = \pi T \sum_{\omega_n > 0} \left ( \frac{2\Delta}{\omega_n} - f_s^+ \right),
	\end{equation}
	and the Usadel equation in the S layer,
	\begin{equation}\label{finUsadelS_x}
	\xi_s^2 \pi T_{cs} \frac{\partial^2 f_s^+}{\partial x^2} - \omega_n f_s^+ + 2 \Delta = 0,
	\end{equation}
	Since the Eq.~\eqref{Usadel_TI_x} can not be solved analytically, to obtain the critical temperature $T_c$ the whole set of equations \eqref{Usadel_TI_re}-\eqref{finUsadelS_x} must be calculated numerically.
	
	\subsection{Single-mode approximation}
	
	In this subsection we present the single mode approximation method.  The solution of the problems \eqref{B1a}-\eqref{finUsadelS} and \eqref{Usadel_TI_re}-\eqref{finUsadelS_x} can be searched in the form of the following anzatz,
	\begin{subequations}\label{Fssingle}
		\begin{equation}
		f_s^+(x,\omega_n)=f(\omega_n) \cos\left(\Omega\frac{x-d_s}{\xi_s}\right),
		\end{equation}
		\begin{equation}
		\Delta(x)=\delta \cos \left(\Omega\frac{x-d_s}{\xi_s}\right),
		\end{equation}
	\end{subequations}
	where $\delta$ and $\Omega$ do not depend on $\omega_n$. The above solution automatically satisfies boundary condition \eqref{Vacuum} at $x=d_s$.

	\subsubsection{Case of $\textbf{h(y)}$}
	
	Substituting expression \eqref{Fssingle} into the \eqref{finUsadelS} we obtain,
	\begin{align}\label{f_om}
	f(\omega_n)=\frac{2 \delta}{ \Omega^2 \pi T_{cs} + \pi T_{cs} \xi_s^2 Q^2 p^2 }.
	\end{align}
	To determine the critical temperature $T_c$ we have to substitute the \Eqs{Fssingle}-\eqref{f_om} into the self-consistency equation \eqref{Delta+} at $T = T_c$. Then it is possible to rewrite the self-consistency equation in the following form,
	\begin{equation}\label{last1}
	\ln \frac{T_{cs}}{T_c} = \psi \left ( \frac{1}{2} + \frac{\Omega^2 + Q^2 p^2}{2}\frac{T_{cs}}{T_c} \right ) - \psi \left ( \frac{1}{2} \right ),
	\end{equation}
	where $\psi$ is the digamma function,
	\begin{equation}\label{digamma}
	\psi(z) \equiv \frac{d}{dz} \ln \Gamma(z), \quad \Gamma(z) = \int_0^{\infty} \eta^{z-1} e^{-\eta} d\eta.
	\end{equation}
	Boundary condition \eqref{B1a} at $x=0$ yields the following equation for $\Omega$,
	\begin{equation}\label{last2}
	\Omega \tan \left ( \Omega \frac{d_s}{\xi_s} \right ) = W^{(p)}(\omega_n).
	\end{equation}
	 Generally, in order to calculate the critical temperature $T_c$, the problem is put on the grid with finite number of the Fourier harmonics $N$ and the following condition should be used,
	\begin{equation}\label{Tc_max}
	T_c=\max \left(T^{(p)}_c\right) \quad p=0,1,2...N.
	\end{equation}
	The critical temperature behavior is found from the solution of the transcendental equations \eqref{last1}, \eqref{last2} as well as \Eq{Tc_max}. Thus, the solution that gives the highest critical temperature $T_c$ is the only one, which is realized physically.

		\begin{figure}[t]
				\centering
				\includegraphics[width=0.7\columnwidth]{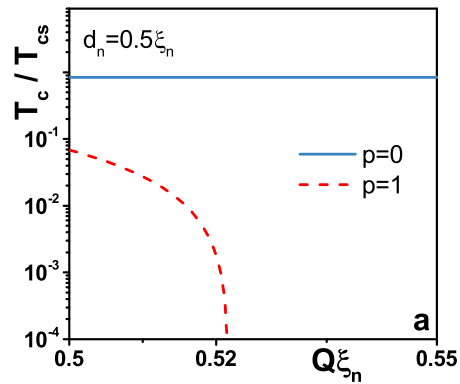}
				\caption
				{(Color online). $T_c(Q)$ dependencies for two harmonic solutions at $ \xi_n h_0/\alpha= 0.1$ . The behavior realized physically is the one which gives the highest $T_c$. The parameters used in the calculations: $\gamma_B=0.1$, $W_f=100 \xi_n$}
				\label{Fourier}
			\end{figure}
			
	However, we find that to calculate the critical temperature it is sufficient to use the zeroth ($p=0$) harmonic of the full Fourier solution for the certain parameter range. In Fig.~\ref{Fourier} we demonstrate the parameter regime, for which the $T_c$ calculation requires consideration of only $p=0$ Fourier component. Such situation is possible due to rapid decay of the $p>0$ components of the solution as functions of $Q$. From the plot it can be noticed that for $Q \xi_n>0.5$ the critical temperature for $p=1$ is not only lower than $T_c$ for $p=0$ but rapidly drops to zero at $Q \xi_n\approx 0.52$.
	
	Such behaviour of the $p>0$ components allows us to operate in the parameter regime by taking appropriate $Q \gg 1$, where the $p=0$ harmonic is sufficient for description of the $T_c$ in the bilayer.
	Since the function oscillates quickly ($Q\xi_n \gg 1$), we also perform averaging of the critical temperature $T_c$ along the $y$ direction.
	
	\subsubsection{Case of $\textbf{h(x)}$} 
	
	As far as the solution of the Eq.~\eqref{Usadel_TI_re} can not be found in analytical form, we calculate the function $f_L$ numerically and solve the problem \eqref{Usadel_TI_re}-\eqref{finUsadelS_x} incorporating single mode approximation \eqref{Fssingle}.
	
	
	\section{Results and Discussion}\label{Results}
	
	In this section we present the results of the critical temperature calculations using the single-mode approximation.
	Some of the parameters are set to the certain values and are not changed throughout the paper, otherwise it is indicated. Such parameters are: $\gamma=0.2$, $\xi_s=\xi_n$ and the width of the junction $W_f=20 \xi_n$.
	%
	
	\subsection{Case of h(y)}
	\begin{figure}[t]
		\centering
		\includegraphics[width=0.95\columnwidth]{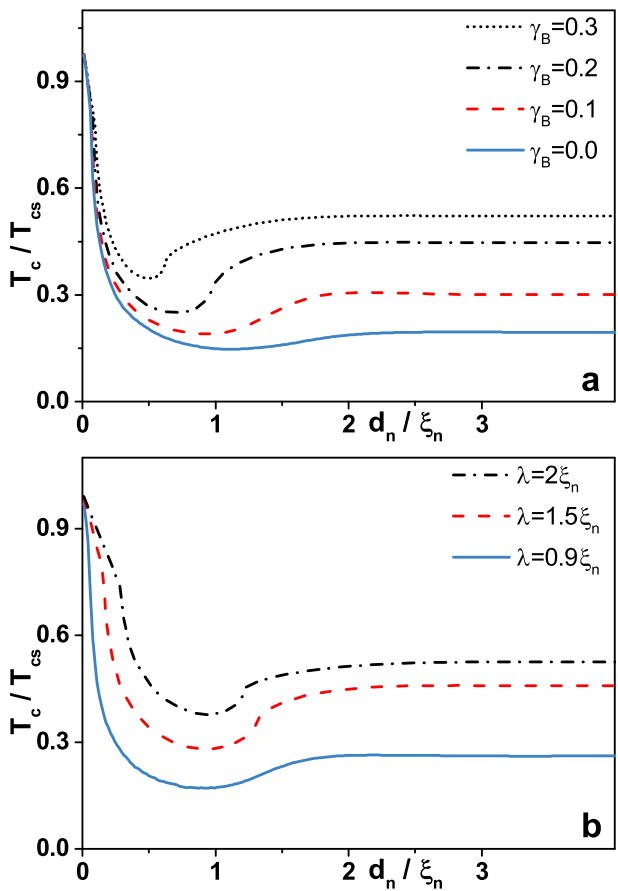}
		\caption
		{(Color online). Behavior of the critical temperature $T_c$ as a function of $d_n$. (a) Each plot corresponds to particular value of the transparency parameter $\gamma_B$: blue solid line to $\gamma_B=0$, red dashed line to $\gamma_B=0.1$ , black dash-dotted line to $\gamma_B=0.2$ and dotted line to $\gamma_B=0.3$. (b) Effect of $\lambda$ on $T_c(d_n)$ dependence. Each curve corresponds to particular value of $\lambda$: blue solid line to $\lambda=0.9\xi_n$, red dashed line to $\lambda=1.5 \xi_n$ and black dash-dotted line to $\lambda=2\xi_n$. The parameters used in the calculations: $ \xi_n h_0 / \alpha =0.25$, $Q=2 \pi/ \lambda$, $\lambda=\xi_n$ (for plot a), $d_s=1.2 \xi_s$}
		\label{Tc2}
	\end{figure}
	In Fig.~\ref{Tc2} (a) the critical temperature dependencies are plotted for different values of the transparency parameter $\gamma_B$. The helical magnetization parameter $ \xi_n h_0/\alpha = 0.25$ and $\lambda=\xi_n$ ($\lambda=2 \pi /Q$). We normalize $T_c$ by its maximum value $T_{cs}$ in the absence of the proximitized TI layer and the TI thickness $d_n$ by the coherence length $\xi_n$. As expected, for perfectly transparent S/TI interface (blue solid line) the critical temperature decreases significantly, showing nonmonotonic behavior with a minimum at $d_n \approx \xi_n$ and eventually saturates at $T_c \approx 0.15 T_{cs}$. For larger values of $\gamma_B$ or at moderate and high resistances of the interface $T_c(d_n)$ saturates at larger temperatures and what is more interesting, the position of the $T_c$ minimum shifts towards smaller values of $d_n$. Unlike $T_c(d_n)$ dependencies in ordinary S/F systems with uniform as well as out of plane spiral magnetization, here the critical temperature does not demonstrate completely reentrant behavior, i. e. the $T_c$ does not vanish in a certain interval of $d_n$. 
	\begin{figure}[t]
		\centering
		\includegraphics[width=\columnwidth]{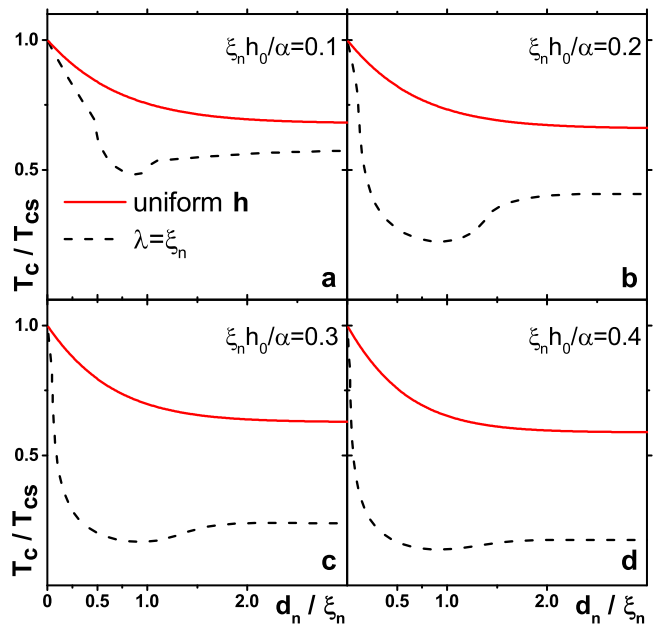}
		\caption
		{(Color online). Comparison of the critical temperature behavior between the S/TI bilayer with uniform magnetization and S/TI bilayer with helical magnetization pattern introduced in Eq.~\eqref{h_y}. The curves were calculated for different values of the $h_0 / \alpha$: plot (a) corresponds to $\xi_n h_0 / \alpha =0.1$, plot (b) to $\xi_n h_0 / \alpha =0.2$, plot (c) to $\xi_n h_0 / \alpha =0.3$ and plot (d) to $\xi_n h_0 / \alpha =0.4$. The general parameters in the bilayers have been set to the identical values such as $\gamma$ and the coherence lengths. The transparency parameter is also same for the both systems $\gamma_B =0.1$. } 
		\label{Tc_comp}
	\end{figure}
	%
	%
	The impact of different $\lambda$ on the critical temperature behavior is depicted in Fig.~\ref{Tc2} (b). Here we took $\gamma_B=0.1$, $\xi_n h_0/\alpha =0.25$ and $d_s=1.2 \xi_s$. From the graph one can notice that $T_c$ becomes more suppressed for smaller values of spatial period $\lambda$ ( which is expressed in terms of $Q$ as $\lambda=2 \pi /Q$), which means that $\lambda$ acts as an additional cause of the superconducting correlations depairing. It is worth mentioning that rather opposite effect has been observed in the S/F hybrid bilayers with out of plane spiral magnetization\cite{Champel2005_2},where $T_c$ experienced enhancement as $Q$ increased. 
	%
	
	\subsection{Uniform and helical magnetization}
	In this subsection we compare the $T_c(d_n)$ behavior in S/TI bilayers with the uniform and helical magnetisation induced on the TI surface. In Fig.~\ref{Tc_comp} the comparison between S/TI with uniform $\textbf{h}$ and with $\textbf{h}(y)$ is shown. From the figure one can see that there is a significant difference in the $T_c(d_n)$ dependence for both cases. First, let us discuss the origin of $T_c$ suppression in the case of uniformly magnetized TI surface. The wavevector of the pair correlations in topological insulator can be written as,
	\begin{equation}
		\kappa_0 = \sqrt{ \frac{2 {\omega_n}}{D} + \frac{4}{\alpha^2}  h_x^2},
	\end{equation}
	where $h_x$ is the magnetization component along the $x$ direction. Here $h_x$ is responsible for depairing of the superconducting correlations and suppresses the critical temperature $T_c$ with the decay length $\xi = 1/\kappa \approx \min \left[ \sqrt{2 \omega_n/ D}, \alpha/2 h_x \right]$. However, $h_y$ component of the magnetization does not play role in suppression of superconducting correlations but introduces a phase shift in the wavefunction, which has no quantitative effect on $T_c$. 
	
	Thus, in Fig.~\ref{Tc_comp} the critical temperature in case of uniform magnetization (red solid lines) expresses monotonic decay due to $h_x$ component. Other type of behavior appears when large enough values of $Q$ are considered in the system. In this case the wavevector acquires additional imaginary term of the form \eqref{f_T} and decay length squared now becomes inverse proportional to $Q$ as $\xi = 1/\kappa \approx \min \left[ \sqrt{2 \omega_n/ D}, \alpha/2 h_0, \sqrt{\alpha / 2 h_0 Q} \right]$.
	%
	%
	In fact, $T_c$ demonstrates nonmonotonic behavior due to fast oscillations of helical magnetization along $y$ axis. This behavior is indicated by black dashed lines (Fig.~\ref{Tc_comp}) and it can be seen that $T_c (d_n)$ loses its nonmonotonicity as $h_0/\xi_n$ grows from clearly pronounced (plots a and b) to hardly recognizable minimum (plots c and d) in the dependence.

	\subsection{Case of h(x)}
\begin{figure}[t]
		\centering
		\includegraphics[width=0.95\columnwidth]{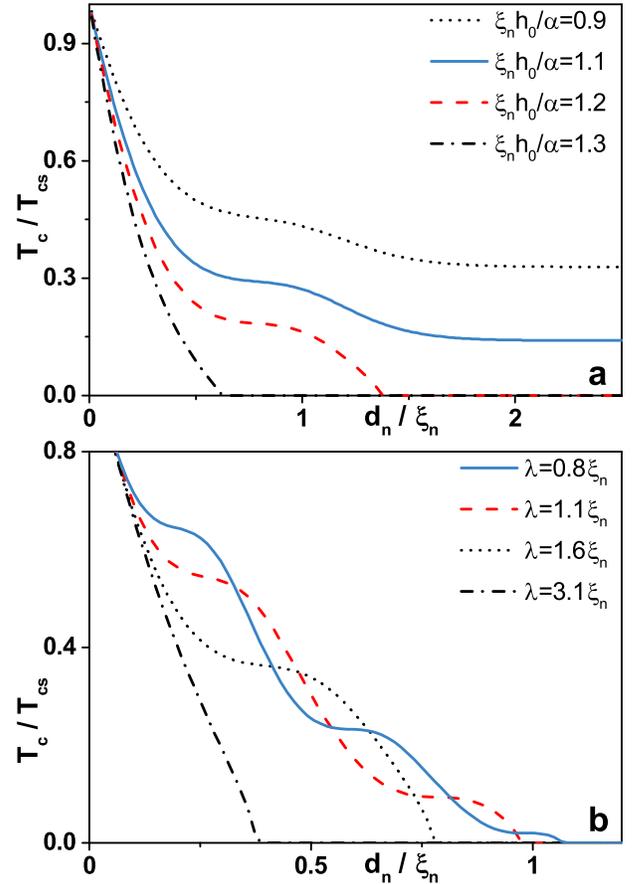}
		\caption
		{(Color online). $T_c(d_n)$ dependencies for the configuration of helical magnetization introduced in Eq.~\eqref{h_x}. (a) Each curve corresponds to particular value of $ h_0/ \alpha$ with fixed helical magnetization parameter $Q=2$. Black dotted line corresponds to $\xi_n h_0 / \alpha =0.9$, blue solid line to $\xi_n h_0 / \alpha =1.1$, red dashed line to $\xi_n h_0 / \alpha =1.2$ and black dash-dotted line to $\xi_n h_0 / \alpha =1.3$ (b) The dependencies correspond to certain values of $\lambda$ and fixed $ \xi_n h_0 / \alpha =1.4$. Blue solid line - $\lambda=0.8\xi_n$, red dashed line - $\lambda=1.1\xi_n$, black dotted line - $\lambda=1.6\xi_n$ and dash-dotted line - $\lambda=3.1\xi_n$The rest of the parameters used in the calculations: $\gamma_B=0$, $d_s=1.2 \xi_s$ }
		\label{Tcx_dn}
	\end{figure}

	Now we turn to the case of S/TI hybrid structure with the TI layer magnetized along the $x$-axis [Fig.~\ref{Fig_Model} (c)]. In Fig.~\ref{Tcx_dn} the critical temperature dependencies as functions of the TI layer thickness $d_n$ are shown. The effect of varying magnetization strength $h_0$ with parameter $Q$ fixed to $Q=2.0$ is shown in the upper plot [Fig.~\ref{Tcx_dn} (a)]. From the plot we can distinguish three types of $T_c$ behavior. For small values of $h_0/\alpha$ the critical temperature demonstrates slightly nonmonotonic behavior with a kink at around $d_n\approx \xi_n$ and eventual saturation (a black dotted line). This nonmonotonic feature becomes more pronounced as $h_0/ \alpha$ is increased (a blue solid line). However, for certain value of magnetization strength $h_0$ the critical temperature drops to zero gradually (a red dashed line). Finally, at relatively large $h_0$ the critical temperature drops sharply down to zero without any bend (a black dash-dotted line).
	
	The origin of such $T_c (d_n)$ curves is a coupling of helical magnetization and momentum of the quasiparticles. However, unlike the magnetization pattern $h(y)$, here the topological insulator TI is magnetized by $h(x)$ along the direction of $d_n$. Hence, the effects on the critical temperature are more explicit and clearer to understand. As it was discussed above, $h_y$ component has no quantitative impact on the magnitude of $T_c$, therefore, the observed effects are purely due to variation of $h_x$ and namely because of its periodicity. Obviously, the number of kinks demonstrated in the Fig.~\ref{Tcx_dn}(a), where we observed just one, depends on magnetization parameter $Q=2\pi/ \lambda$. In Fig.~\ref{Tcx_dn}(b) the critical temperature behavior for different $Q$ is shown. It can be seen that the smaller spatial magnetization period $\lambda$ the more kinks are produced in the $T_c$.
	\begin{figure}[t]
		\centering
		\includegraphics[width=\columnwidth]{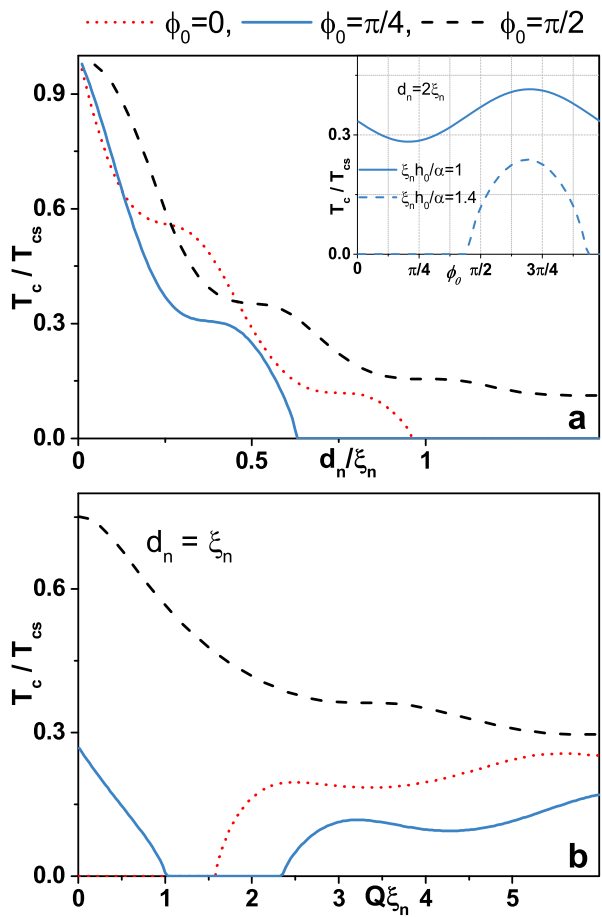}
		\caption
		{(Color online). Influence of the arbitrary initial phase $\phi_0$ in the magnetization pattern $\textbf{h}(x)=h_0(\cos (Qx + \phi_0), \sin (Qx + \phi_0), 0)$. Each curve corresponds to particular value of $\phi_0$: red dotted line to $\phi_0=0$, blue solid line to $\phi_0=\pi / 4$ and black dashed line to $\phi_0=\pi/2$. (a) $T_c(d_n)$ dependencies calculated for $ \xi_n h_0/ \alpha=1.4$ and $ \lambda= \xi_n$. The inset plot shows $T_c$ behavior as a function of phase $\phi_0$ for fixed thickness $d_n=2\xi_n$ and two different values of $\xi_n h_0/ \alpha=1, 1.4$. (b) $T_c(Q)$ curves calculated for $d_n =\xi_n$ and $ \xi_n h_0/ \alpha=1.2$. The parameters used in the calculations:$\gamma_B=0$, $ \xi_n h_0 / \alpha =0.25$ and $d_s=1.2 \xi_s$ }
		\label{Tcx_phi0}
	\end{figure}
	In the calculations above we assumed that the magnetization pattern $\textbf{h}(x)$ at $x=0$ reduces to $\textbf{h}(0)=h_0(1, 0, 0)$, which implies that the initial `phase' is $0$. In practice, it may be possible to have arbitrary initial phase in the experimental samples. It is very important to consider such possibility since $T_c$ decays significantly in our system as a function of the TI thickness $d_n$. We can take into account $\phi_0$ simply by rewriting the magnetization pattern \eqref{h_x} as,
	\begin{equation}
	\textbf{h}(x)=h_0(\cos (Qx + \phi_0), \sin (Qx + \phi_0), 0).
	\end{equation}

	In Fig.~\ref{Tcx_phi0} (a) the effect of various initial $\phi_0$ on $T_c(d_n)$ for fixed $h_0/\alpha =1.4$ and $\lambda=\xi_n$ is illustrated. From the plot we observe that while $\phi_0=0$ and $\phi_0=\pi/4$ contribute to faster decay of $T_c$ as a function of $d_n$ (red dotted and blue solid line), the critical temperature has higher values at almost every $d_n$ for $\phi_0=\pi/2$ (black dashed line). The inset shows $T_c$ as a function of $\phi_0$ for fixed $d_n=2\xi_n$. 
	
	Another interesting result can be noticed in Fig.~\ref{Tcx_phi0} (b) illustrating $T_c(Q)$ dependencies for the same values of $\phi_0$ and fixed TI layer thickness $d_n=\xi_n$. One can recognize that depending on $\phi_0$ the critical temperature behaves differently as $Q$ changes. For $\phi_0=0$ (red dotted line) there is no superconductivity in the $Q\xi_n$ interval $\left[0, 1.5\right]$ since $T_c$ is completely suppressed by slowly evolving near extremum $h_x$ magnetization component at the vicinity of the S/TI interface.  However, for $\phi_0=\pi/4$ (blue solid line) $T_c$ decays rapidly and vanishes at $Q\xi_n\approx 1$ but eventually restores at $Q\xi_n\approx 2.4$. Finally in the case of $\phi_0=\pi/2$ (black dashed line) the critical temperature is almost not suppressed at small values of $Q$, but decays gradually as $Q$ is further increased.
	
	\section{Conclusion}\label{Conclusion}
In this work we have formulated a theoretical approach and presented the results of a quantitative investigation of the superconducting critical temperature in the S/TI hybrid structure, where an in-plane helical magnetization is induced at the TI surface. 
The calculations are based on the quasiclassical Usadel equations, taking into account SOC at the surface of the topological insulator. We have found that in the case of in-plane helical magnetization $\textbf{h}(y)$, evolving along the interface, the calculations reveal nonmonotonic behavior of the critical temperature as a function of the TI layer thickness with a well pronounced minimum, the effect which is absent in the case of uniform magnetization. Moreover, in the case of helical magnetization, evolving perpendicular to the interface $\textbf{h}(x)$, the critical temperature demonstrates highly nonmonotonic behavior as well. However, this dependence has been shown to be qualitatively different from the case of $\textbf{h}(y)$, showing number of kinks, which depends on helical magnetization parameters.

The results are important for further understanding of the underlying physics and potential future applications of superconductor - TI hybrid systems. 
	
	\begin{acknowledgments}
		 T.K. and A.S.V. acknowledge support of the Mirror Laboratories project of the HSE University and the Bashkir State Pedagogical University. V.S.S. acknowledges support of the joint French (ANR) / Russian (RSF) Grant ``CrysTop" (20-42-09033). A.A.G. acknowledges support by the European Union H2020-WIDESPREAD-05-2017-Twinning project SPINTECH under Grant Agreement No. 810144.
		 
	\end{acknowledgments}


\begin{thebibliography}{99}
		\addcontentsline{toc}{chapter}{Bibliography}
		
		
\bibitem{Fu2007}
L. Fu, C. L. Kane and E. J. Mele Phys. Rev. Lett. \textbf{98}, 106803 (2007).

\bibitem{Hasan2010}
M. Z. Hasan and C. L. Kane, Rev. Mod. Phys., \textbf{82}, 3045 (2010).

\bibitem{Sato2017}
M. Sato and Y. Ando, Reports on Progress in Physics \textbf{80}, 076501 (2017).

\bibitem{book1}
S.-Q. Shen, \textit{Topological Insulators Dirac Equation in
Condensed Matters} (Berlin: Springer, 2012).

\bibitem{book2}
G. Tkachov \textit{Topological Insulators: The Physics of Spin Helicity in Quantum Transport} (Singapore: Pan Stanford, 2015).


\bibitem{Sarma2006}
S. D. Sarma, M. Freedman and C. Nayak, Phys. Today \textbf{59}, 32 (2006).

\bibitem{Aguado2020}
R. Aguado and L. P. Kouwenhoven, Phys. Today \textbf{73}, 44 (2020).

		
\bibitem{Qi2011}
X.-L. Qi and  S.-C. Zhang, Rev. Mod. Phys. \textbf{83}, 1057 (2011).
	

\bibitem{Fu2008} 
L. Fu, and C. L. Kane, Phys. Rev. Lett. \textbf{100}, 096407 (2008).

\bibitem{Tanaka2009}
Y. Tanaka,  T. Yokoyama, and N. Nagaosa, Phys. Rev. Lett. \textbf{103}, 107002 (2009).

\bibitem{Sato2009}
M. Sato and S. Fujimoto, Phys. Rev. B \textbf{79}, 094504 (2009).

\bibitem{Alicea2012} 
J. Alicea, Rep. Prog. Phys. \textbf{75}, 076501 (2012).
		
\bibitem{Beenakker2013}
C. W. J. Beenakker, Annu. Rev. Condens. Matter Phys. \textbf{4}, 113 (2013).

\bibitem{Tkachov2013}
G. Tkachov and E. N. Hankiewicz, Phys. Rev. B \textbf{88}, 075401 (2013).


\bibitem{Demler1997}
E. A. Demler, G. B. Arnold, and M. R. Beasley, Phys. Rev. B \textbf{55}, 15174 (1997).
		
\bibitem{Ozaeta2012R}
A. Ozaeta, A. S. Vasenko, F. W. J. Hekking, and F. S. Bergeret, Phys. Rev. B \textbf{86}, 060509(R) (2012).

\bibitem{Bergeret2013}
F. S. Bergeret and I. V. Tokatly, Phys. Rev. Lett. \textbf{110}, 117003 (2013).

\bibitem{Bobkova2017}
I. V. Bobkova and A. M. Bobkov, Phys. Rev. B \textbf{95}, 184518 (2017).

\bibitem{BergeretRMP}
F. S. Bergeret, A. F. Volkov, and K. B. Efetov, Rev. Mod. Phys. \textbf{77},
1321 (2005).

\bibitem{BuzdinRMP}
A. I. Buzdin, Rev. Mod. Phys. \textbf{77}, 935 (2005).

		
\bibitem{Fominov2002}
Y. V. Fominov, N. M. Chtchelkatchev, and A. A. Golubov, Phys. Rev. B \textbf{66}, 014507 (2002).

\bibitem{Fominov2010}
Ya. V. Fominov, A. A. Golubov, T. Yu. Karminskaya, M. Yu. Kupriyanov, R. G. Deminov, L. R. Tagirov, JETP Lett. \textbf{91}, 308 (2010) [Pis'ma ZhETF \textbf{91}, 329 (2010)].


\bibitem{Buzdin1982}
A. I. Buzdin, L. N. Bulaevskii, and S. V. Panyukov, JETP Lett. \textbf{35}, 178 (1982) [Pis'ma ZhETF \textbf{35}, 147 (1982)].
		
\bibitem{Buzdin1991}
A. I. Buzdin and M. Yu. Kupriyanov, JETP Lett. \textbf{53}, 321 (1991) [Pis'ma ZhETF \textbf{53}, 308 (1991)].
		
\bibitem{Ryazanov2001}
V. V. Ryazanov, V. A. Oboznov, A. Yu. Rusanov, A. V. Veretennikov, A. A. Golubov, and J. Aarts, Phys. Rev. Lett. \textbf{86}, 2427 (2001).

\bibitem{Oboznov2006}
V. A. Oboznov, V. V. Bol'ginov, A. K. Feofanov, V. V. Ryazanov, and A. I. Buzdin, Phys. Rev. Lett. \textbf{96}, 197003 (2006).
		
\bibitem{Vedyayev2006}
A. V. Vedyayev, N. V. Ryzhanova, N. G. Pugach, Journal of Magnetism and Magnetic Materials, \textbf{305}, 53 (2006).
		
\bibitem{Vasenko2008}
A. S. Vasenko, A. A. Golubov, M. Yu. Kupriyanov, and M. Weides, Phys. Rev. B \textbf{77}, 134507 (2008).
		
\bibitem{Bakurskiy2017}
S. V. Bakurskiy, V. I. Filippov, V. I. Ruzhickiy, N. V. Klenov, I. I. Soloviev, M. Yu. Kupriyanov, A. A. Golubov, Phys. Rev. B \textbf{95}, 094522 (2017).


\bibitem{Kontos2001}
T. Kontos, M. Aprili, J. Lesueur, and X. Grison, Phys. Rev. Lett. \textbf{86}, 304 (2001).
		
\bibitem{Vasenko2011}
A. S. Vasenko, S. Kawabata, A. A. Golubov, M. Yu. Kupriyanov, C. Lacroix, F. S. Bergeret, and F. W. J. Hekking, Phys. Rev. B \textbf{84}, 024524 (2011).


\bibitem{Jiang1995}
J. S. Jiang, D. Davidovi\`{c}, D. H. Reich, and C. L. Chien, Phys. Rev. Lett. \textbf{74}, 314 (1995).
		
\bibitem{Tagirov1998}
L. R. Tagirov, Physica C \textbf{307}, 145 (1998).
		
\bibitem{Proshin2001}
Yu. N. Proshin, Yu. A. Izyumov, and M. G. Khusainov, Phys. Rev. B \textbf{64}, 064522 (2001).
		
\bibitem{Khaydukov2018}	
Yu. N. Khaydukov, A. S. Vasenko, E. A. Kravtsov, V. V. Progliado, V. D. Zhaketov, A. Csik, Yu. V. Nikitenko, A. V. Petrenko, T. Keller, A. A. Golubov, M. Yu. Kupriyanov, V. V. Ustinov, V. L. Aksenov, and B. Keimer, Phys. Rev. B \textbf{97}, 144511 (2018).

\bibitem{Karabassov2019}
T. Karabassov, V. S. Stolyarov, A. A. Golubov, V. M. Silkin, V. M. Bayazitov, B. G. Lvov,
and A. S. Vasenko, Phys. Rev. B \textbf{100}, 104502 (2019).


\bibitem{Zuyzin2016}
A. Zyuzin, M. Alidoust, and D. Loss, Phys. Rev. B \textbf{93}, 214502 (2016).

		
\bibitem{Bergeret2014}
F.S. Bergeret and I. V. Tokatly, Phys. Rev. B \textbf{89}, 134517 (2014).
		
\bibitem{Jacobsen2015}
S. H. Jacobsen and J. Linder, Phys. Rev. B \textbf{92}, 024501 (2015).
		
\bibitem{Bujnowski2019}
B. Bujnowski, R. Biele, and F.S. Bergeret, Phys. Rev. B \textbf{100}, 224518 (2019).
		
\bibitem{Eskilt2019}
J. R. Eskilt, M. Amundsen, N. Banerjee, and Jacob Linder, Phys. Rev. B \textbf{100}, 224519 (2019).
		
\bibitem{Nashaat2019}
M. Nashaat, I. V. Bobkova, A.M. Bobkov, Y.M. Shukrinov, I.R. Rahmonov, and K. Sengupta, (2019).

\bibitem{Bobkova2016}
I. V. Bobkova, A. M. Bobkov, A. A. Zyuzin, and M. Alidoust, Phys. Rev. B \textbf{94}, 134506 (2016).

\bibitem{Alidoust2018}
M. Alidoust, Phys. Rev. B \textbf{98}, 245418 (2018).

\bibitem{Alidoust2020}
M. Alidoust, Phys. Rev. B \textbf{101}, 155123 (2020).

\bibitem{Lu2019}
Y. Lu and T.T. Heikkilä, Phys. Rev. B \textbf{100}, 104514 (2019).



\bibitem{Alidoust2017}
M. Alidoust and H. Hamzehpour, Phys. Rev. B \textbf{96}, 165422 (2017).
		 

\bibitem{Champel2005}
T. Champel and M. Eschrig, Phys. Rev. B \textbf{71}, R220506 (2005).
		
\bibitem{Champel2005_2}
T. Champel and M. Eschrig, Phys. Rev. B \textbf{72}, 054523 (2005).

\bibitem{Champel2008}
T. Champel, T. Löfwander, and M. Eschrig, Phys. Rev. Lett. \textbf{100}, 077003 (2008).

\bibitem{Pugach2017}
N.G. Pugach, M. Safonchik, T. Champel, M.E. Zhitomirsky, E. Lähderanta, M. Eschrig, and C. Lacroix, Appl. Phys. Lett. \textbf{111}, 162601 (2017).

\bibitem{Pugach2017_2}
N.G. Pugach and M.O. Safonchik, JETP Lett. \textbf{107}, 302 (2018).

\bibitem{Pugach2018}
N.G. Pugach, M.O. Safonchik, D.M. Heim, and V.O. Yagovtsev, Phys. Solid State \textbf{60}, 2237 (2018).


\bibitem{Bode2007}
M. Bode, M. Heide, K. Von Bergmann, P. Ferriani, S. Heinze, G. Bihlmayer, A. Kubetzka, O. Pietzsch, S. Blügel, and R. Wiesendanger, Nature \textbf{447}, 190 (2007).


\bibitem{Brun2016}
C. Brun, T. Cren, and D. Roditchev,  Supercond. Sci. Technol. \textbf{30} 013003 (2017).

		

\bibitem{Belzig1999}
W. Belzig, F. K. Wilhelm, C. Bruder, G. Sch\"{o}n, and A. D. Zaikin, Superlattices Microstruct. \textbf{25}, 1251 (1999).

\bibitem{Usadel1970}
K. D. Usadel, Phys. Rev. Lett. \textbf{25}, 507 (1970).
		
\bibitem{KL}
M. Yu. Kuprianov and V. F. Lukichev, JETP \textbf{67}, 1163 (1988) [ZhETF \textbf{94}, 139 (1988)].
		
\bibitem{gamma_b}
E. V. Bezuglyi, A. S. Vasenko, V. S. Shumeiko, and G. Wendin, Phys. Rev. B \textbf{72}, 014501 (2005); E. V. Bezuglyi, A. S. Vasenko,
E. N. Bratus, V. S. Shumeiko, and G. Wendin, ibid. \textbf{73}, 220506(R) (2006).
		
\bibitem{GKLO}
A. A. Golubov, M. Yu. Kupriyanov, V. F. Lukichev, and A. A. Orlikovskii, Sov. J. Microelectron. \textbf{12}, 191 (1984) [Mikroelektronika \textbf{12}, 355 (1983)].


		


		
	\end{thebibliography}
\end{document}